\documentclass{article}
\usepackage[nonatbib,preprint]{neurips_2021}

\usepackage{microtype}
\usepackage{graphicx}
\usepackage{subcaption}
\usepackage{caption}
\usepackage{booktabs} %
\usepackage{xcolor}
\usepackage{amsthm}
\usepackage{amssymb}
\usepackage{amsmath}
\usepackage{mathtools}
\usepackage{verbatim}
\usepackage[backend=biber]{biblatex}
\usepackage{bm}
\usepackage{bbm}
\usepackage{algorithm}
\usepackage{algpseudocode}
\usepackage{wrapfig}
\DeclareMathOperator{\rgt}{rgt}
\usepackage{hyperref}
\usepackage{cleveref}

\newcommand{\michael}[1]{{\color{blue}{\footnotesize\sf[MJC: #1]}}}

\title{Learning Revenue-Maximizing Auctions With Differentiable Matching}
\author{%
  Michael J. Curry \\
  University of Maryland\\
  \texttt{curry@cs.umd.edu} \\
\And
  Uro Lyi \\
  University of Maryland\\
  \texttt{ulyi@umd.edu} \\
 \And
 Tom Goldstein \\
 University of Maryland \\
 \texttt{tomg@cs.umd.edu} \\
 \And
  John P. Dickerson \\
  University of Maryland\\
  \texttt{john@cs.umd.edu} \\
}

\begin{document}
\maketitle

\begin{abstract}
We propose a new architecture to approximately learn incentive compatible, revenue-maximizing auctions from sampled valuations. Our architecture uses the Sinkhorn algorithm to perform a differentiable bipartite matching which allows the network to learn strategyproof revenue-maximizing mechanisms in settings not learnable by the previous RegretNet architecture. In particular, our architecture is able to learn mechanisms in settings without free disposal where each bidder must be allocated exactly some number of items. In experiments, we show our approach successfully recovers multiple known optimal mechanisms and high-revenue, low-regret mechanisms in larger settings where the optimal mechanism is unknown.

\end{abstract}

\section{Introduction}\label{sec:intro}

Auctions have been held for millennia, since at least Classical antiquity~\cite{krishna2009auction}, and have played an important complementary role to set-price sales and bargaining to exchange goods.  In recent decades, the advent of computation has resulted in a surge of large-scale fielded auctions in a variety of important industries, such as broadcasting~\cite{Leyton17:Economics}, advertising~\cite{Edelman07:Internet}, electricity markets~\cite{cramton2017electricity}, and many others~\cite{Milgrom17:Discovering,Roth18:Marketplaces}.  These developments have made auctions not only of theoretical interest, but also of great practical importance.

Auction \emph{design} is the problem faced by an auctioneer who, given uncertain knowledge of the demands of auction participants, wishes to set the rules of the auction to ensure, via proper incentive structure, a desirable outcome.  
In the usual theoretical model of auctions~\cite{Parsons11:Auctions}, the bidders have some private valuation of the items, but the distribution from which these valuations are drawn is common knowledge. The auctioneer solicits bids from participants, awards the items up for sale to the winners, and charges some amount of money to each.  Bidders may, however, choose to strategically lie about their valuation given their knowledge of the auction rules and anticipated behavior of other participants, resulting in a Bayes-Nash equilibrium which may be very hard for the auction designer to predict.

To avoid this problem, an auctioneer may simply wish to design a strategyproof (or truthful) mechanism in which participants are incentivized to truthfully reveal their valuations. The distribution of bids is then simply the valuation distribution itself, so it becomes easy to predict the results of the auction in expectation. While satisfying the strategyproofness constraint, the auctioneer can additionally optimize total utility, their own revenue, or other desirable properties.

If the auctioneer's goal is to maximize the welfare of all participants while maintaining strategyproofness, the Vickrey-Clarke-Groves (VCG) mechanism always gives a solution \cite{Vickrey61:Counterspeculation,Clarke71:Multipart,Groves73:Incentives}. By contrast, if the auctioneer wishes to maximize revenue, strategyproof mechanisms are much harder to find. Some revenue-maximizing strategyproof mechanisms are known in limited cases: when selling a single item, the Myerson auction is strategyproof and maximizes revenue \cite{myerson1981optimal}, and for selling multiple items to one agent, some results are known \cite{DBLP:conf/sigecom/DaskalakisDT15,kash2016optimal,pavlov2011optimal,manelli2006bundling}. But even for the simple case of selling \emph{two} items to \emph{two} agents, there has been almost no progress, with the major exception of a breakthrough for the special case where items take on at most two discrete valuations \cite{yao2017dominant}. 

The theoretical difficulty of devising revenue-maximizing strategyproof auctions has resulted in a number of attempts to approximate them. Recently, \cite{DBLP:conf/icml/Duetting0NPR19} presented a method, RegretNet, for learning approximately incentive compatible mechanisms given samples from the bidder valuations. They parameterize the auction as a neural network, and learn to maximize revenue, and maintain strategyproofness, by gradient descent. This learning approach to auction design can replicate some known-optimal auctions, and can learn good auctions in settings where the optimal auction is not known. It has been extended in a variety of ways \cite{Curry20:Certifying,Tacchetti19:Neural,golowich2018deep,Kuo20:Proportionnet,feng2018deep,Rahme21:Auction,rahme2020permutation,Peri21:PreferenceNet}.

 The RegretNet architecture has architectural features to ensure that no item is over-allocated, and that each bidder receives the correct amount of goods. This is accomplished through a combination of softmax and minimum operations, depending on the exact bidder demand constraints. 
 
 However, this approach only works when some ad-hoc combination of these operations is sufficient to enforce the constraints. This is not always possible. Consider the case where the auctioneer needs to ensure that every participant receives exactly $k$ items. This corresponds, for instance, to an offline version of the ``Display Ads'' problem \cite{Feldman2009-nm}, without free disposal. Such equality constraints cannot be enforced using the min-of-softmax approach.
 
 We observe that an auction allocation for bidders whose utilities are linear functions of their valuations amounts to a matching assigning items to bidders. Using this observation, we present an alternative approach which explicitly applies matching constraints on the output, using the Sinkhorn algorithm \cite{cuturi2013sinkhorn} to solve a discrete matching problem as part of an end-to-end differentiable architecture.

This approach is among the first attempts to use the techniques of differentiable optimization \cite{Agrawal2019:cvxpylayers} for learned mechanism design -- concurrent work \cite{liu2021neural} uses a differentiable sorting operator to learn generalized second-price auctions. By simply modifying the constraints in our matching problem, our proposed approach can produce feasible allocations without changes to the overall network architecture for many different bidder demand constraints. Our approach successfully recovers known optimal mechanisms in multiple settings including settings (such as exactly-one-demand auctions) in which RegretNet would be unable to guarantee that its output allocations adhere to the constraints.

\section{Related Work}\label{sec:rw}

\paragraph{Auction Design and Learning}

Auction mechanisms are functions from bids to allocations and payments; if one assumes sample access to the valuation distribution, it is natural to treat auction design as a learning problem, and there has been much work in this area.

One thread of work has been learning-theoretic, determining the sample complexity for various known families of auctions to estimate properties like revenue~\cite{balcan2016sample,cole2014sample,morgenstern2016learning} or incentive compatibility \cite{Balcan19:Estimating}.  %

Another thread of work, sometimes called ``differentiable economics'', represents auction mechanisms using general parametric function approximators, and attempts to optimize them using gradient descent. \cite{DBLP:conf/icml/Duetting0NPR19} introduce several neural network architectures to find revenue-maximizing auctions, including RochetNet, which works for single-agent auctions and is strategyproof by constructions, and RegretNet, which represents auctions as a general neural network and includes a term in the loss function to enforce strategyproofness.

Further work has built directly on both RochetNet and RegretNet \cite{Kuo20:Proportionnet,Curry20:Certifying,Rahme21:Auction,rahme2020permutation}. Others have taken a similar approach but applied to different mechanism design problems, including finding welfare-maximizing auctions \cite{Tacchetti19:Neural} and facility location \cite{golowich2018deep}. Other applications of gradient-based methods to problems in mechanism design include \cite{heidekruger2021equilibrium,weissteiner2020deep}.

\paragraph{Single-agent auction learning} The case of selling multiple items to a single agent is reasonably well understood. \cite{rochet1987necessary} gives a characterization of strategyproof single-agent mechanisms -- their utility functions must be monotone and convex. Based on this characterization, \cite{DBLP:conf/icml/Duetting0NPR19} design RochetNet, a network architecture for single-agent mechanism learning which is guaranteed to be perfectly strategyproof. \cite{shen2019automated} also provide network architectures for the single-agent setting which can be made strategyproof by construction. The authors of both works are able to learn mechanisms and then prove them optimal using the theoretical framework of \cite{DBLP:conf/sigecom/DaskalakisDT15}.

We wish to explicitly contrast these architectures with our own approach. In a single-agent setting, they work better than a RegretNet-style architecture, and as mentioned can always guarantee perfect strategyproofness. In this work, however, we focus on general architectures which can work in both single- and multi-agent settings.

\paragraph{Optimal Transport and the Sinkhorn Algorithm}

Optimal transport \cite{kantorovich1942translocation} is the problem of moving one set of masses to another mass while minimizing some cost function. In its infinite dimensional form, where it amounts to finding the cost-minimizing joint distribution between two marginal continuous probability distributions, it has wide applications in pure mathematics \cite{villani2003topics} as well as machine learning \cite{Arjovsky17:WGAN,genevay2018learning}. Interestingly, \cite{DBLP:conf/sigecom/DaskalakisDT15,kash2016optimal} use the mathematical tools of infinite-dimensional optimal transport in the context of mechanism design theory. 

Our work here is not directly related. By contrast, we the discrete form of optimal transport, which is essentially a formulation of minimum cost bipartite matching (in our case, between agents and items). This discrete problem can be numerically solved in a number of ways, see \cite{peyre2019computational} for an overview. In particular, we focus on the Sinkhorn algorithm \cite{cuturi2013sinkhorn}: a fast, GPU-parallelizable iterative method for solving the entropy-regularized version of the optimal transport problem, which can be used as a  differentiable bipartite matching operation \cite{mena2018learning,grover2019stochastic,tay2020sparse,emami2018learning,cuturi2019ranking}. We use the Sinkhorn algorithm to compute matchings between agents and items.

\paragraph{Differentiable optimization and deep learning}
Recently, there has been broad interest in mixing convex optimization problems with deep learning. For many convex optimization problems, the derivative of the optimal solution with respect to parameters of the objective or constraints is well defined, so it is possible to define a neural network layer that will output a feasible, optimal solution to some optimization problem. This is useful if one wants to use optimization with deep learning in a ``predict and optimize'' pipeline \cite{ferber2020mipaal,wilder2019melding}, to enforce that network outputs satisfy some constraints, or if interesting operations can be formulated in terms of optimization problems \cite{gould2019deep}.

One family of approaches \cite[e.g.,][]{Agrawal2019:cvxpylayers,amos2017optnet} involves solving the convex optimization problem using standard solvers, then using the implicit function theorem to compute the gradient for the backward pass. In a contrasting approach, when using an iterative method to solve the optimization problem, it is also possible to use automatic differentiation to simply backpropagate through the numerical operations performed by the solver. We employ this latter approach with the aforementioned Sinkhorn algorithm in order to compute feasible matchings in a differentiable manner

\section{Differentiable Economics and Combinatorial Optimization}\label{sec:nn}

\paragraph{General auction setting} We consider an auction setting with $n$ bidders and $m$ items. Each bidder $i$ has a private valuation function, $v_{i}: [0, 1]^m \rightarrow \mathbb{R}_{\geq 0}$, that maps any subset of the items to a real number.
We assume that $v_i$ is drawn from a known distribution $F_i$, but the realized valuation $v_i$ is private and unknown to the auctioneer. Each bidder $i$ then submits their bids $b_i \in \mathbb{R}^m$. Let $b = (b_1, b_2, ..., b_n)$ be the bids from all bidder. The auction mechanism is then a combination of an allocation mechanism and payment mechanism, $(g(b), p(b))$. The $g$ outputs an allocation $(g_1, g_2, ..., g_n) $ where $g_{ij}$ is the probability of allocating item $j$ to bidder $i$. The payment mechanism $p$ outputs $(p_1, p_2, ..., p_n)$ where $p_i$ is how much bidder $i$ is charged. Each bidder then receives some utility $u_i(v_i, b) = v_i(g_i(b)) - p_i(b)$.

\paragraph{Quasilinear utilities, strategyproofness, and individual rationality} Allowing a distinct valuation for every combination of items may result in a combinatorial explosion. A simplifying assumption is that a bidder may have a single valuation for each item. Then their valuation $v_i$ is simply a vector in $\mathbb{R}^m$, and their utility is simply $u_i(v_i, b) = \langle v_i, g_i(b) \rangle - p_i(b)$.

Since a bidder's true valuation is private they are free to strategically report bids to maximize their utility. Thus, a we require our mechanism to be \textbf{strategyproof} or \textbf{dominant-strategy incentive compatible (DSIC)}, meaning each bidder's utility is maximized by reporting truthfully i.e. $u_i(v_i, (v_i, b_{-i})) \geq u_i(v_i, (v_i^\prime, b_{-i})$ for any other $v_i^\prime$ where $b_{-i}$ are all the bids except the $i$th bidder. \textbf{Regret} is defined by the following equation: $\text{rgt}_i(v_i, b) = \max_{v_i^\prime} u_i(v_i, (v_i^\prime, b_{-i})) - u_i(v_i, b)$,
and represents the utility the bidder could have gained from lying in their bid. We can then also say that when a mechanism is DSIC, the regret for truthful bidding for bidder $i$ should be 0 for every player $i$ with truthful valuation $v_i$, for any opponent bids $b_{-i}$. 

Another desirable property for an auction for it to be (ex post) \textbf{individually rational}. This means each bidder receives a non-negative utility from the auction i.e. $u_i(v_i, (v_i, b_{-i})) \geq 0$ for all bidders $i$, all possible valuations $v_i$, and all other bids $b_{-i}$. Without this guarantee bidders might choose not to participate in the auction at all because they are concerned about being left worse off than they were before the auction.

\subsection{Bidder Demand Types}
Here we define common bidder demand constraints which are used in the experiments.

The value of a bundle of goods for a \textbf{unit-demand} bidder is equal to the maximum value of a single item in the bundle: $v_i(S) = \max_{j \in S} v_i(j)$.
In this sort of auction, there is no need to consider allocations of more than one item to each bidder since those provide that same utility as only allocating the most desirable item within that bundle to that bidder. Thus, by restricting allocations to allocate at most one item per bidder, we can again treat bidder utility as quasilinear. More generally, there can be $k$-unit-demand auctions where a bidder's value of a bundle of goods is the sum of the top $k$ items in the bundle~\cite{Zhang20:Learning}.  %

We also consider \textbf{exactly-one demand}, where each bidder must be assigned exactly one item. In this setting, the bidder cannot be assigned more or fewer than one item and receives the value of the assigned item. 
Additionally, this demand-type can be extended to a more general exactly-$k$ demand setting where each bidder must be allocated $k$ items.

In all these settings, in general allocations might be randomized so that bidders can receive items with different probabilities. This corresponds to allowing fractional allocations, so that in the unit-demand and exactly-one-demand settings, the one item the bidder receives can be a mixture of fractions of multiple items.

\subsection{RegretNet}
The RegretNet architecture consists of two neural networks, the allocation network (denote it $g$) which outputs a matrix representing the allocation of each item to each agent and payment network (denote it $p$) which outputs the payments for each bidder~\cite{DBLP:conf/icml/Duetting0NPR19}.

RegretNet guarantees individual rationality by making the payment network use a sigmoid activation, outputting a value $\tilde{p} \in [0,1]$. The final payment is then $\left(\sum_j v_{ij} g_{ij}\right)\tilde{p}$, ensuring that the utility of each bidder is non-negative.

\cite{DBLP:conf/icml/Duetting0NPR19} present network architectures to learn under additive, unit-demand and combinatorial valuations. Their architecture utilizes a traditional feed-forward neural network with modifications in the output layer to ensure a valid allocation matrix. In the additive setting, the allocation probabilities for each item must be at most one: this is enforced by taking a row-wise softmax on the network outputs. For unit-demand auctions, each bidder wants at most one item, so the network takes a row-wise softmax of one matrix and a column-wise softmax of another to ensure item allocation is less than one. The final output is then the elementwise min of these two, ensuring that both item allocation and unit-demand constraints are enforced.

For training, the RegretNet architecture uses gradient descent on an augmented Lagrangian which includes terms to maximize payment while also containing terms to enforce the constraint that regret should be zero:
{\small \begin{equation}
    \mathcal{L}_{\bm \lambda}(\bm v) = -\sum_i p_i(\bm v) + \sum_i \lambda_i \widehat{\rgt}_i(\bm v) + \frac{\rho}{2} \sum_i \left(\widehat{\rgt}_i(\bm v)\right)^2
\label{eq:regretnetloss}
\end{equation}}
Here $\widehat{\rgt}_i$ denotes an empirical estimate of regret produced by running gradient ascent on player $i$'s portion of the network input to maximize their utility, computing their possible benefit from strategically lying. During training, the Lagrange multipliers $\lambda$ are gradually maximized, incentivizing the minimization of regret.

\section{Bipartite Matching for Auctions}\label{sec:bipartite}
The auction allocation problem is equivalent to finding a minimum-cost bipartite matching between bidders and items, for some cost matrix. We solve this bipartite matching problem by formulating it as an optimal transport problem. Using the Sinkhorn algorithm, we are able to solve this optimal transport problem in an end-to-end differentiable way, yielding a valid matching that our network can use to learn an optimal allocation mechanism.

\begin{figure*}
\centering
\includegraphics[width=0.7\textwidth]{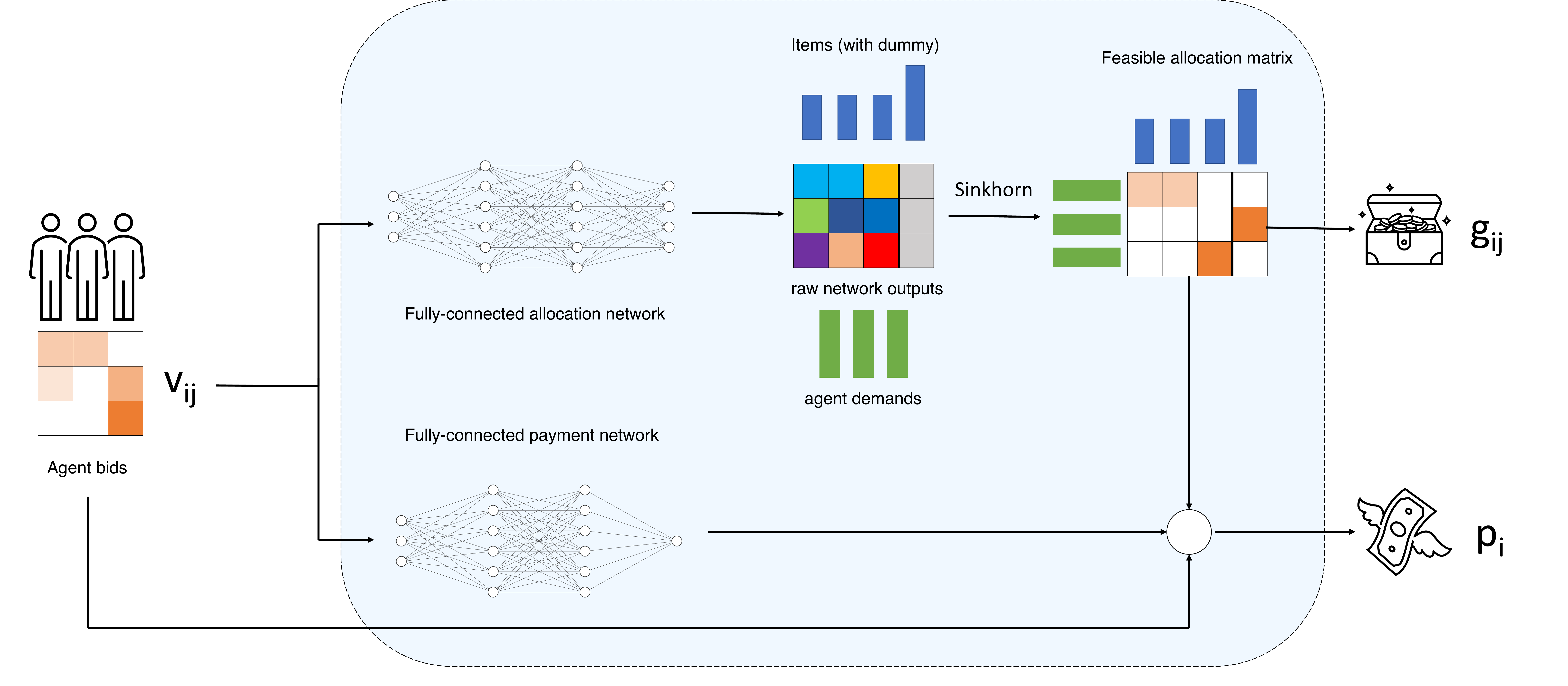}
\caption{A schematic showing the Sinkhorn-based mechanism network. The agents' bids (here, assumed to be truthful) on each item act as input to a feedforward network, whose output is used as the cost matrix $\bm C_{ij}$ for the Sinkhorn algorithm. Marginals, specified separately, ensure that the Sinkhorn output is a feasible allocation. The payment network charges a fraction of the value of items that each agent wins.}
\label{fig:sinkhorndiagram}
\end{figure*}
\subsection{Matching Linear Program}
For marginal vectors $\bm a$ and $\bm b$ -- here representing constraints for agents and items respectively, each with a dummy component that represents no item or no agent -- the optimal transport problem is as follows.
\begin{equation}
    \min_{\bm P} \sum_{i=1}^{n+1} \sum_{j=1}^{m+1} \bm P_{ij} \bm C_{ij} \text{ s.t. }\, \bm P \mathbbm{1}_{m+1} = \bm a,\,\mathbbm{1}_{n+1}^T \bm P  = \bm b
    \label{eq:discreteot}
\end{equation}
By specifying $\bm a$ and $\bm b$, we can specify the demand constraints of the problem:
\begin{itemize}
    \item $k$-demand: $\bm a_i = k$, $\bm a_{n+1} = m$, $\bm b_j = 1$, $\bm b_{m+1} = kn$
    \item exactly-$k$-demand: $\bm a_i = k$, $\bm a_{n+1} = m - kn$, $\bm b_j = 1$, $\bm b_{m+1} = 0$
\end{itemize}

Here the $n+1$ and $m+1$ values of the marginals represent the dummy agent and item respectively. More complex constraints, such as different $k$ for different agents, can also be accommodated. Also, note in the exactly-$k$-demand settings there must be at least $kn$ items, so $m \geq kn$. An optimal solution to the matching LP will always be a permutation matrix \cite{birkhoff1946tres}, or more generally on one of the extreme points of the constraint set.

\subsection{Entropy-Regularized Optimal Transport and the Sinkhorn Algorithm}

The problem in \Cref{eq:discreteot} can be solved as a linear program. However, \cite{cuturi2013sinkhorn} observed that by adding an entropy penalty to the problem, it can be solved in a practical and scalable way using the matrix-scaling Sinkhorn algorithm. The new objective of the problem is
\begin{equation*}
    \min_{\bm P}\sum_{i=1}^N\sum_{j=1}^M \bm P_{ij} \bm C_{ij} + \epsilon \sum_{i=1}^N \sum_{j=1}^M \bm P_{ij} \log \bm P_{ij}
\end{equation*}

Given marginals and a cost matrix, \cite{cuturi2013sinkhorn} provides an iterative algorithm to solve this problem. It can be implemented in a variety of ways; we use the following stabilized log-domain updates \cite{peyre2019computational}:
\begin{equation*}
    \bm f_i = -\epsilon \log\left(\sum_{j} \exp\left(-\frac{C_{ij} - \bm g_{j}}{\epsilon}\right)\right) + \epsilon \log \bm a,\,
    \bm g_j = -\epsilon \log\left(\sum_{i} \exp\left(-\frac{C_{ij} - \bm f_{i}}{\epsilon}\right)\right) + \epsilon \log \bm b 
\end{equation*}

where $\bm P_{i,j} = e^{\bm f_i/\epsilon}e^{-C_{ij}/\epsilon}e^{\bm g_j/\epsilon}$.
In addition to enabling the use of the iterative algorithm, the regularization term makes the problem strongly convex. Thus, the derivative of the optimal solution with respect to the cost matrix is well-defined. To approximate this derivative, we choose to unroll several iterations of the Sinkhorn updates (which consist only of differentiable operations) and use automatic differentiation.

The $\epsilon$ parameter controls the strength of the regularization. If it is large the output $\bm P$ will be nearly uniform; as it goes to zero we recover the original problem and $\bm P$ will be a permutation matrix. For our purposes, we set $\epsilon$ large enough to avoid vanishing gradients, but small enough that the output allocations are still nearly permutation matrices. 

\subsection{Network Architecture}

Our architecture, like RegretNet, uses a pair of networks to compute allocations and payments. We denote the allocation and payment networks $g^w$ and $p^w$ respectively, where $w$ are the network weights.

 The main distinction from RegretNet is in the allocation mechanism. Like RegretNet we utilize a traditional feedforward neural network that takes all given bids as input. We use the output of the network as the cost matrix $\bm C_{ij}$ to the Sinkhorn algorithm (if there are dummy agents or items, they receive cost 0).
 Marginals are specified separately depending on the problem setting.
 
 Finally, the output of the Sinkhorn algorithm (after truncating the row and column for dummy variables) provides the final probabilities $g^w_{ij}$ of allocating item $j$ to bidder $i$. 

As in RegretNet, the final payment for bidder $i$ is $\left( \sum_j v_{ij} g_{ij} \right) \tilde{p}^w_i$ where $\tilde{p}^w_i \in [0, 1]$ is the $i$th output from a feedforward payment network. This ensures that the mechanism is individually rational as the bidders will never be charged more than their utility gained from the allocated items.

\subsection{Settings That RegretNet Cannot Represent}

RegretNet provides three different architectures depending on bidder demand type and valuation functions. For its unit-demand and combinatorial architectures, it employs a ``min-of-softmax'' approach to ensure both that agents receive at most 1 item or bundle, and items are not overallocated.

We particularly emphasize the case of exactly-$k$ demand, which describes the situation where the auction designer is obligated to ensure that every participant receives $k$ items, no matter what they bid. This captures, for instance, an offline version of the classic Display Ads setting \cite{Feldman2009-nm}, without free disposal, where every ad buyer is contractually guaranteed a certain number of ad slots. For multiple agents and multiple items, RegretNet cannot represent this demand type via the min-of-softmax approach.

(In some special cases, ad-hoc changes could be made to the original RegretNet architecture to support settings beyond additive and unit-demand. For example, to enforce ``at-most-$k$'' demand, one can just scale the unit-demand outputs by a factor of $k$. In the special single-agent case, it is possible to produce exactly-$k$ demand by removing the dummy item which allows for the possibility of allocating no item.)

By contrast, in multi-agent settings with exactly-$k$ demand, the Sinkhorn-based architecture can ensure feasible outputs by simply specifying the correct marginals. Our network architecture is also independent of the demand type since only the marginals change rather than the entire network output structure.

\subsection{Optimization and Training}
Like RegretNet, our training objective is \Cref{eq:regretnetloss}. For each batch we estimate the empirical regret, $\widehat{\rgt}_i(w)$ by performing gradient ascent for each bidder on the network inputs, approximating a misreport for each bidder.

In our experiments, the network for both the payment and allocation network had 2 hidden layers of 128 nodes with Tanh activation functions, optimized using the Adam optimizer \cite{kingma2014adam} with a learning rate of $10^{-3}$. The training set consisted of $2^{19} = 524,288$ valuation profiles. We used batch sizes of 4,096 and ran a total of 100 epochs. For each batch, we computed the best misreport using $25$ loops of gradient ascent with a learning rate $0.1$. We also incremented $\rho$ every two batches and and updated the Lagrange multipliers, $\lambda$, every 100 batches. Experiments were all run on single 2080Ti GPUs, on either a compute node or workstation with 32GB RAM, using PyTorch \cite{paszke2019pytorch}.

For evaluation, we used 1,000 testing examples and optimized the misreports of these examples for 1,000 iterations with learning rate $0.1$. For each of the 1,000 samples we use ten random initialization points for the misreport optimization and use the maximum regret from these. At both train and test time we used the Sinkhorn algorithm with an $\epsilon$ parameter of $0.05$, and a stopping criterion of a relative tolerance of $0.01$.  Additionally we used an $\epsilon$ schedule (as suggested by \cite{schmitzer2019stabilized,cuturi2019ranking}) of 10 steps from $1$ down to the final value of $0.05$. Pseudocode for computing the allocation network output is given in algorithm \ref{alg:allocate}.

\begin{algorithm}[tb]
\small
\caption{Sinkhorn allocation procedure}
\label{alg:allocate}
\begin{algorithmic}
\State {\bfseries Input:} Bid $\bm v \in \mathbb{R}^{m \times n}$, allocation network output $f(\bm v; \theta)$, marginals $\bm a \in \mathbb{R}^{n+1}$ and $\bm b \in \mathbb{R}^{m+1}$, Sinkhorn epsilon schedule $\epsilon_1, \cdots, \epsilon_T$, tolerance $t$
\State {\bfseries Output:} Feasible allocation matrix $\bm g_{ij} \in \mathbb{R}^{m \times n}$
\State {\bfseries Initialize:} $\bm f = 0^{n+1}, \bm g = 0^{m+1}$, $\bm C = f(\bm v; \theta)$
\For{$\epsilon$ in $\epsilon_0, \cdots, \epsilon_T$}
\State $\bm P_{ij} = e^{\bm f_i/\epsilon}e^{-\bm C_{ij}/\epsilon}e^{\bm g_j/\epsilon}$
\While{$\max_i |\sum_j \bm P_{ij} - \bm a_i|/\bm a_i \geq t$}
\State $\bm f_i = -\epsilon \log\left(\sum_{j} \exp\left(-\frac{\bm C_{ij} - \bm g_{j}}{\epsilon}\right)\right) + \epsilon \log \bm a$,\, $\bm g_j = -\epsilon \log\left(\sum_{i} \exp\left(-\frac{\bm C_{ij} - \bm f_{i}}{\epsilon}\right)\right) + \epsilon \log \bm b$
\State $\bm P_{ij} = e^{\bm f_i/\epsilon}e^{-\bm C_{ij}/\epsilon}e^{\bm g_j/\epsilon}$
\EndWhile
\EndFor
\State $\bm g_{ij} = \bm P_{ij}$ for $i=1..n, j=1..m$
\end{algorithmic}
\end{algorithm}

\section{Experiments}\label{sec:exp}

\subsection{Optimal single-agent mechanisms}

We start by recovering two known optimal mechanisms in the single bidder case. We emphasize that if the only goal were to learn single-agent auctions, architectures such as RochetNet \cite{DBLP:conf/icml/Duetting0NPR19} or that of \cite{shen2019automated} would work better. Our architecture works for both single-agent and multi-agent auctions. Following previous work \cite{DBLP:conf/icml/Duetting0NPR19,rahme2020permutation,Rahme21:Auction}, we use these single-agent experiments as a test case to make sure it recovers some known optimal mechanisms, before continuing on to multi-agent settings where optimal auctions are not known and where RochetNet or \cite{shen2019automated} cannot work.

\paragraph{Unit-demand}

\begin{figure}
\centering
    \begin{subfigure}[h]{0.4\textwidth}
        \includegraphics[width=\textwidth]{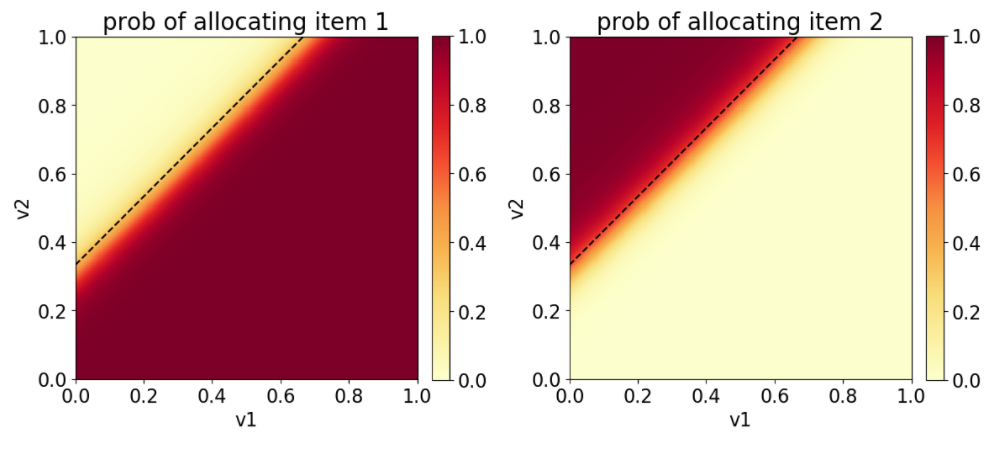}
        \caption{Exactly-One}
        \label{fig:single_exact}
    \end{subfigure}
    \begin{subfigure}[h]{0.4\textwidth}
        \includegraphics[width=\textwidth]{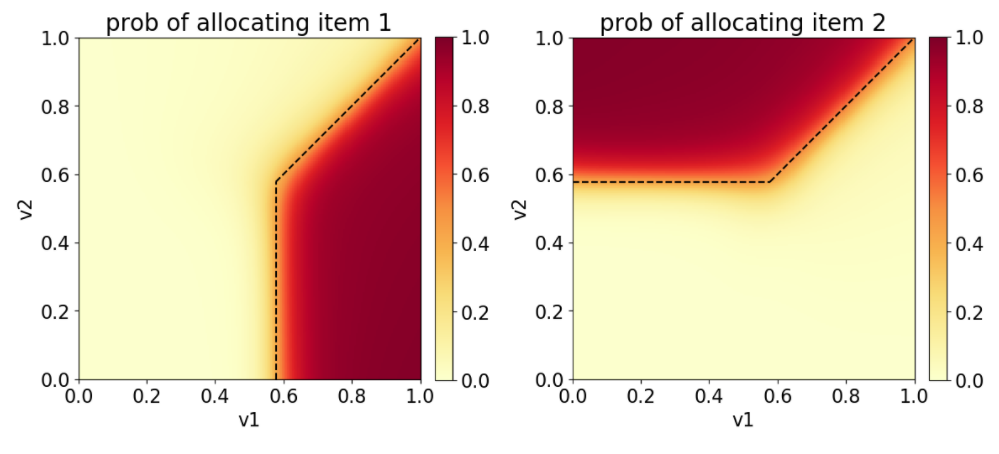}
        \caption{Unit-Demand}
        \label{fig:single_unit}
    \end{subfigure}
    \caption{Heatmaps of learned mechanisms for single-bidder two-item with $v_1, v_2 \sim U[0, 1]$ are shown in (a)-(b). 
    The optimal mechanism boundaries are the black dotted lines. 
    }
    \label{fig:single_allocs}
\end{figure}

We first recover the optimal mechanism in the unit-demand single-agent two-item setting with item values drawn from $U[0, 1]$. The optimal mechanism, also approximately recovered in \cite{DBLP:conf/icml/Duetting0NPR19}, is from \cite{pavlov2011optimal}: it is to offer each good for a price of $\frac{\sqrt{3}}{3}$. The learned allocation probabilities are shown in Figure \ref{fig:single_unit} with the boundary of the optimal analytic mechanism denoted by a dotted line. The x and y axis are the valuation of the bidder for item one and two respectively. The color represents the allocation probability output by the learned mechanism with the darker color corresponding to higher probability. Quantitatively, the learned mechanism has small regret and slightly higher than the optimal mechanism likely due to the small amount of regret. 

\paragraph{Exactly-one demand}
The second optimal mechanism was a deterministic mechanism in the same single agent, two-item setting with valuations on $U[0,1]$. However, this time the agent will be allocated exactly one item (instead of at most one). \cite{kash2016optimal} shows that the optimal mechanism is to offer one of the items for free or the other for a price of $\frac{1}{3}$.

The boundary of this optimal mechanism is shown in Figure \ref{fig:single_exact} as the black dotted line. 
The revenue in Figure \ref{fig:single_table} is slightly higher than the optimal, again likely due to the presence of small regret.

\begin{table}
    \centering
    \begin{tabular}{ @{}ccccc@{} } 
    \toprule
    Setting & Rev & Regret & Opt Rev & Train Time\\
    \midrule
    Unit-demand & 0.397 (0.248) & $<0.001$ ($<0.001$) & 0.393 & 3h25m  \\
    RegretNet Unit-Demand & 0.381 (0.258) & $<0.001$ ($<0.001$) & 0.393 & 18m34s \\
    Exactly-one & 0.079 (0.127) & $0.001$ (0.001) & 0.069 & 1h21m \\
    \bottomrule
    \end{tabular}
    \caption{Table of mean revenue and regret from learned mechanism in test set alongside revenue of optimal mechanism for 1 agent, 2 items, with item valuations distributed independently on $U[0,1]$.}
    \label{fig:single_table}
\end{table}
\subsection{Multi-agent setting}
\begin{figure}
    \centering
    \begin{subfigure}[b]{0.47\textwidth}
    \centering
    \includegraphics[width=\columnwidth]{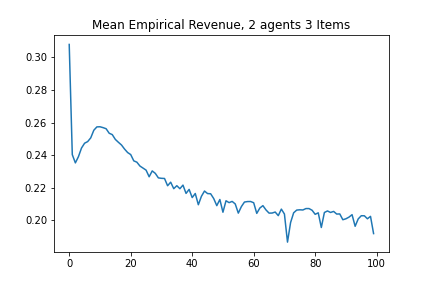}
    \caption{Mean revenue for exactly-one demand, 2 agent 3 item setting}
    \label{fig:meanrev}
\end{subfigure}
\hfill
\begin{subfigure}[b]{0.47\textwidth}
    \centering
    \includegraphics[width=\columnwidth]{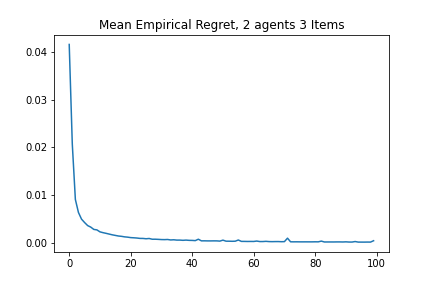}
    \caption{Mean empirical regret for exactly-one demand, 2 agent 3 item setting}
    \label{fig:meanrgt}
\end{subfigure}
\end{figure}

We now experiment in settings where the optimal strategyproof mechanism is unknown. Specifically, we study 2 bidders and either 3 or 15 items where valuations for each item are drawn from $U[0,1]$. 
A typical analytic baseline is a separate Myerson auction for each item, but this only works in additive settings. Instead, as a baseline, we compute the revenue from an \cite{Vickrey61:Counterspeculation,Clarke71:Multipart,Groves73:Incentives}. The VCG auction is strategyproof, and maximizes welfare rather than revenue. It works by charging each bidder the harm they cause to other bidders, which in the settings we test may be very small or even close to zero. 
Table \ref{tab:twoagenttable} contains the revenues for both mechanism (unsurprisingly higher for the learned auction) as well as the average observed regret on the testing sample, while figures \ref{fig:meanrev} and \ref{fig:meanrgt} show training plots for the two agent, 3 item, exactly-one demand case.

\subsection{Comparison To RegretNet}

RegretNet is capable of representing unit-demand auctions, but not the exactly-one setting. We compare performance to it in this case; results are shown in Table \ref{tab:twoagenttable}. (Runtimes are significantly faster than those reported in \cite{DBLP:conf/icml/Duetting0NPR19} likely due to our much larger batch sizes.) For the one-agent, two item setting, both architectures approximate the optimal mechanism so performance is similar. Performance is also similar for the 2 agent, 3 item setting. In the 15-item setting, revenue with the Sinkhorn architecture is somewhat smaller. In all cases the computational cost of RegretNet is significantly lower, as it requires only two softmax operations instead of many iterations of the Sinkhorn algorithm.

\begin{table}
    \centering
    \begin{tabular}{ @{}ccccccc@{} } 
    \toprule
    Setting & Agents & Items & Rev & Regret & VCG Rev & Time  \\
    \midrule
    Unit & 2 & 3  & .876 (.322) & .001 (.001) & .048 & 6h20m\\
    RegretNet Unit & 2 & 3 & .878 (.337) & $<.001$ ($<.001$) & .048 & 20m28s \\
    Unit & 2 & 15 & .886 (.113) & .001 ($<.001$) & .002 & 3h10m  \\
    RegretNet Unit & 2 & 15 & .999 (.123) & $<.001$ ($<.001$) & .002 & 25m45s\\
    \midrule
    Exactly-1 & 2 & 3 & .194 (.064) & .004 (.008) & .049 & 2h38m\\
    RegretNet Exactly-1 & 2 & 3 & N/A & N/A & .049 & N/A  \\
    Exactly-1 & 2 & 15 & .571 (.030) & .003 (.015) & .002 & 1h29m\\
    RegretNet Exactly-1 & 2 & 15 & N/A & N/A & .002  & N/A  \\
    \bottomrule
    \end{tabular}
    \caption{Table of mean revenue and regret from learned mechanism in test set along with revenue from VCG auction, for 2 agents with valuations distributed independently on $U[0,1]$}
    \label{tab:twoagenttable}
\end{table}

\section{Strengths, Limitations, and Potential Impacts}
Structurally, our architecture has the benefit of remaining the same for any quasi-linear bidder utilities and constraints, with only the Sinkhorn marginals changing. The use of the Sinkhorn algorithm allows us to enforce these constraints explicitly and even tackle new settings, such as exactly-$k$ demand, that the existing RegretNet architecture would be unable to handle. 

Our Sinkhorn-based architecture has a higher computational cost than the comparable RegretNet architecture, as the Sinkhorn algorithm requires many iterations. The Sinkhorn algorithm also requires two additional hyperparameters, $\epsilon$ and a tolerance before the algorithm terminates.

We find that the $\epsilon$ parameter plays a crucial role in the training of the algorithm. If the $\epsilon$ is too small, training fails, likely due to vanishing gradients. However, with too large of an $\epsilon$ the mechanism becomes too smooth and unable to approximate the sharp boundaries that tend to show up in optimal mechanisms leading to suboptimal revenue. (For further discussion of $\epsilon$, see Appendix \ref{app:sinkhornepsilon} in the supplemental material.)
Our architecture has an inductive bias towards deterministic allocations. This may pose a limitation where revenue-maximizing mechanisms are nondeterministic, but it may be an advantage if they are deterministic or if determinism is desirable.
By setting $\epsilon$ small enough it is possible to ensure near-determinism in allocations. However, because the mechanisms are trained under a higher $\epsilon$, there are regions where the price charged becomes too high, increasing regret. Because training directly with small $\epsilon$ is difficult, we cannot directly guarantee that we train deterministic mechanisms.

Following previous work that uses neural networks to approximate optimal auctions (including but not limited to \cite{DBLP:conf/icml/Duetting0NPR19,shen2019automated,rahme2020permutation,Rahme21:Auction}), we train on synthetic data. While in principle deep-learning-based methods could be used to train on truthful bids from real-world bidders, and doing so would be extremely interesting, such data is hard to come by. We see learned auctions as, at least in part, a tool for pushing theory forward, so using synthetic data remains interesting when it is drawn from valuation distributions for which analytic solutions have remained out of reach. 

There are general ethical implications of revenue-maximizing auction design, in terms of potential impact on bidders and society as a whole. For instance, making it easier for sellers to extract revenue from bidders might be viewed as bad (if the bidders are thought of as ordinary consumers) or good (if the auctioneer is selling a public resource to private entities). However, since our work is still closely to the theoretical models discussed above, we do not see any direct ethical implications -- it is very unlikely that deep-learned auctions will be directly deployed in the near future.

\section{Conclusion and Future Work}

We have presented a new architecture for learned auctions which uses the Sinkhorn algorithm to perform a differentiable matching operation to compute an allocation. Our architecture works for a variety of bidder demand constraints by encoding them into the marginals used in the Sinkhorn algorithm. This new architecture allows the network to guarantee valid allocations in settings where RegretNet could not.

We show that our approach successfully recovers optimal mechanisms in settings where optimal mechanisms are known, and achieves good revenue and low regret in larger settings. Future work might include extending the Sinkhorn architecture to more directly support randomized allocations, further computational improvements, or further extensions to other mechanism design problems where the allocation decision can be expressed as a matching. 

\noindent\textbf{Acknowledgments.}
Curry and Dickerson were supported in part by NSF CAREER Award IIS-1846237, NSF D-ISN Award \#2039862, NSF Award CCF-1852352, NIH R01 Award NLM-013039-01, NIST MSE Award \#20126334, DARPA GARD \#HR00112020007, DoD WHS Award \#HQ003420F0035, and a Google Faculty Research Award.  Goldstein was supported by the ONR MURI program, the AFOSR MURI Program, the National Science Foundation (DMS-1912866), the JP Morgan Faculty Award, and the Sloan Foundation.
We thank Ahmed Abdelkader, Kevin Kuo, and Neehar Peri for comments on earlier drafts of this work, and Jason Hartline for helpful commentary.

\printbibliography

\clearpage
\appendix

\section{Effect of Sinkhorn Epsilon}
\label{app:sinkhornepsilon}

\begin{figure}[h]
\centering
    \begin{subfigure}[h]{0.4\textwidth}
        \includegraphics[width=\textwidth]{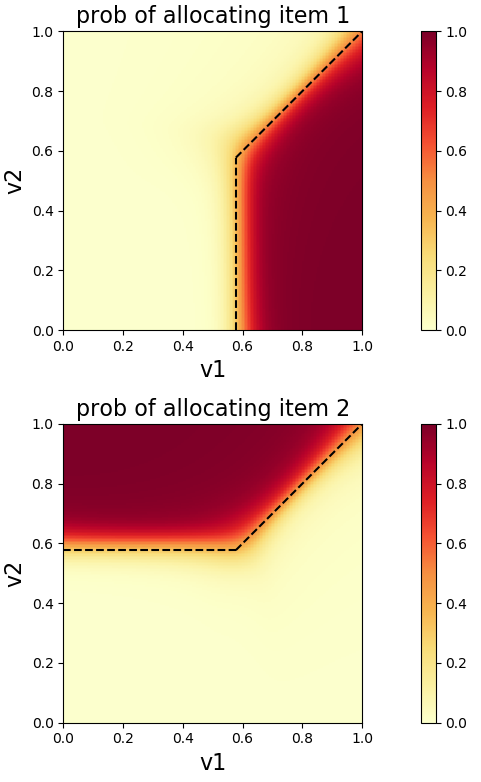}
        \caption{$\epsilon = 0.05$}
        \label{fig:high_temp}
    \end{subfigure}
    \begin{subfigure}[h]{0.4\textwidth}
        \includegraphics[width=\textwidth]{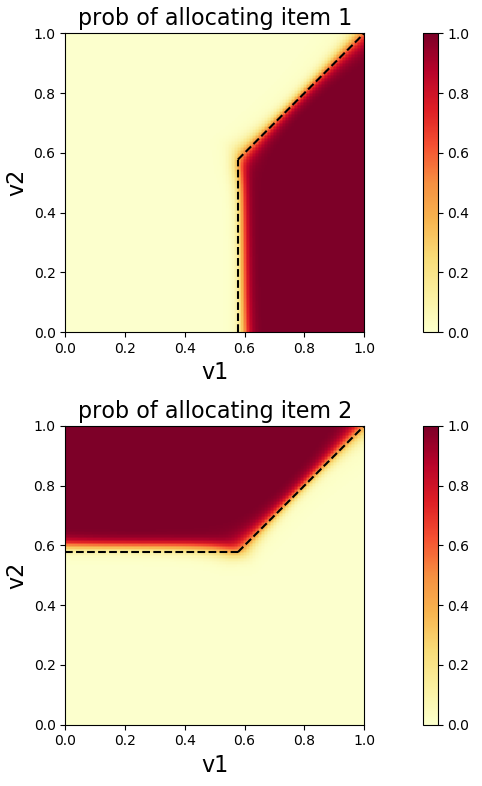}
        \caption{$\epsilon = 0.02$}
        \label{fig:low_temp}
    \end{subfigure}
    \caption{The following figures illustrate the effect of Sinkhorn temperature on allocation mechanism (c) higher Sinkhorn $\epsilon$ value and (d) lower Sinkhorn $\epsilon$ value. }
    \label{fig:sinkhornepsilon}
\end{figure}

\paragraph{Effect of Sinkhorn $\epsilon$} 
Figures \ref{fig:low_temp} and \ref{fig:high_temp} highlights the effect of the Sinkhorn parameters on the final mechanism. The lower the $\epsilon$ in the Sinkhorn algorithm the sharper the boundary becomes, as with less entropy regularization, the optimal matching is closer to deterministic. However, we found that very small values of epsilon led to problems during training; the learned mechanism would choose to never allocate items, likely due to vanishing gradients. One can make an almost-perfectly-deterministic mechanism by decreasing $\epsilon$ at test time, but this can increase regret as the learned payments no longer agree with the allocations.

\end{document}